\pdfoutput=1

\documentclass[%
 aps,
 pra,
 floatfix,
 amsmath, amssymb,
 reprint,%
]{revtex4-1}

\usepackage[utf8]{inputenc}
\usepackage[T1]{fontenc}
\usepackage{hyperref}

\usepackage{graphicx}
\usepackage{dcolumn}
\usepackage{bm}
\usepackage{braket} 
\usepackage{siunitx}

\usepackage{xcolor}

\begin{document}

\preprint{APS/123-QED}

\title[Atom tomography]{Spatial tomography of individual atoms in a quantum gas microscope}


\author{Ottó Elíasson}
\thanks{Ottó Elíasson and Jens S. Laustsen contributed equally to this work}
 \affiliation{ 
Department of Physics and Astronomy, Aarhus University, 8000 Aarhus C, Denmark
}%
\author{Jens S.\ Laustsen}
\thanks{Ottó Elíasson and Jens S. Laustsen contributed equally to this work}
 \affiliation{ 
Department of Physics and Astronomy, Aarhus University, 8000 Aarhus C, Denmark
}%

\author{Robert Heck}
 \affiliation{ 
Department of Physics and Astronomy, Aarhus University, 8000 Aarhus C, Denmark
}%

\author{Romain Müller}
 \affiliation{ 
Department of Physics and Astronomy, Aarhus University, 8000 Aarhus C, Denmark
}%

\author{Carrie A. Weidner}
 \affiliation{ 
Department of Physics and Astronomy, Aarhus University, 8000 Aarhus C, Denmark
}%

\author{Jan J.\ Arlt}
 \affiliation{ 
Department of Physics and Astronomy, Aarhus University, 8000 Aarhus C, Denmark
}%

\author{Jacob F.\ Sherson}%
 \email{sherson@phys.au.dk}
\affiliation{ 
Department of Physics and Astronomy, Aarhus University, 8000 Aarhus C, Denmark
}%


\date{\today}

\begin{abstract}
We demonstrate a method to determine the position of single atoms in a three-dimensional optical lattice. Atoms are sparsely loaded from a far-off-resonant optical tweezer into a few vertical planes of a cubic optical lattice positioned near a high-resolution microscope objective. In a single realization of the experiment, we pin the atoms in deep lattices and then acquire multiple fluorescence images with single-site resolution. The objective is translated between images, bringing different lattice planes of the lattice into focus. The applicability of our method is assessed using simulated fluorescence images, where the atomic filling fraction in the lattice is varied.
This opens up the possibility of extending the domain of quantum simulation using quantum gas microscopes from two to three dimensions.
\end{abstract}

\pacs{Valid PACS appear here}
\maketitle

\section{Introduction}

It is becoming increasingly clear that a complete understanding of quantum materials requires an understanding of phenomena like correlations, fluctuations and entanglement down to the single-particle level. Neutral atoms are among the prime candidates for performing scalable quantum simulations. Within that category there are currently two main platforms that allow for measurements of quantum properties down to the level of an individual atom.

The first of these relies on the usage of multiple tightly focused optical tweezers directly loaded with atoms from a magneto-optical trap~\cite{Barredo2016,endresAtombyatomAssemblyDefectfree2016}. As a result, the atoms are typically not in the motional ground state, and the inter-atomic couplings are mediated by Rydberg interactions. Tweezer-trapped atoms can, however, be cooled to the ground state~\cite{serwaneDeterministicPreparationTunable2011, kaufmanCoolingSingleAtom2012}, and this enables tunnel-coupling between adjacent sites~\cite{Kaufman2014,murmannTwoFermionsDouble2015}. Large arrays of ground-state cooled single atoms have recently been realized, demonstrating the scalability of these systems~\cite{norciaMicroscopicControlDetection2018,cooperAlkalineEarthAtomsOptical2018}.

A second platform studies ultracold atoms loaded into the ground state of wavelength-scale optical lattices by means of high-resolution microscopy in so-called quantum gas microscopes (QGMs)~\cite{Bakr2009,shersonSingleatomresolvedFluorescenceImaging2010}. Atoms in adjacent lattice sites are typically closer together than the corresponding resolution limit of the imaging system,  which renders their distinction challenging, but possible~\cite{karskiNearestNeighborDetectionAtoms2009}. Both of these platforms offer a high degree of control and tunability~\cite{lewensteinUltracoldAtomsOptical2012}. The work presented here combines aspects of both methods.

\begin{figure}[tp]
  \centering
    \includegraphics[width=1\columnwidth]{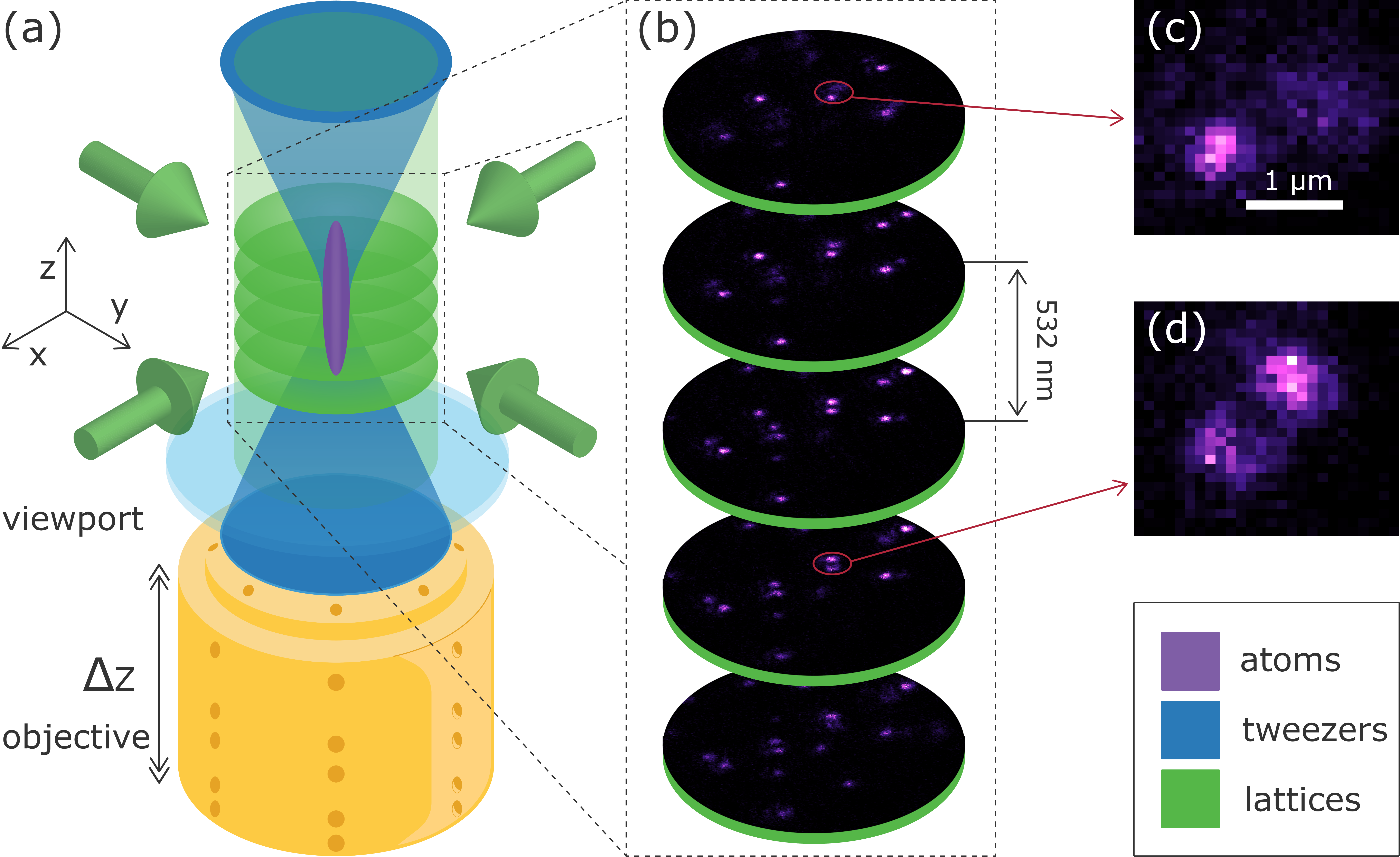} 
  \caption{The experimental configuration. (a) A small BEC (violet) is created in a 
  DMD-generated optical tweezer potential (blue) projected through a $0.69\,$NA microscope objective.  An optical lattice (green) is created along the vertical ($z$) direction by retro-reflection off of a viewport (light blue). The direction of the two horizontally-propagating lattice beams is indicated by the green arrows. The tweezer-trapped atoms are released into the vertical lattice, where they spread out before being pinned in a deep cubic optical lattice and their fluorescence is imaged. The objective (yellow) can be shifted between exposures, allowing for consecutive images of single atoms at different focal positions (b). Figures (c) and (d) show images of two atoms loaded in different planes from the same experimental realization, illustrating that one atom moves into focus while the other is defocused as the objective moves.} 
  \label{fig:One}
\end{figure}

Quantum gas microscopes operate in a regime where a short lattice spacing allows for tunnel coupling between atoms in adjacent potential wells. These systems have proven to be especially useful for quantum simulation of many-body systems~\cite{grossQuantumSimulationsUltracold2017}. Prominent applications in bosonic systems include the study of 1D Heisenberg spin-chains~\cite{Fukuhara2013}, quantum random walks~\cite{preissStronglyCorrelatedQuantum2015}, the links between mixed quantum states and thermal ensembles~\cite{kaufmanQuantumThermalizationEntanglement2016}, and the realization of many-body localized states~\cite{choiExploringManybodyLocalization2016}. One of the biggest advantages of QGMs is direct access to high-order particle-particle correlations, which are essential for the description of many-body quantum states. The construction of fermionic QGMs~\cite{cheukQuantumGasMicroscopeFermionic2015,parsonsSiteResolvedImagingFermionic2015a,hallerSingleatomImagingFermions2015} enabled the study of quantum magnetism which relies heavily on such correlations. Measurements of correlations both in the spin and charge sectors have revealed long-range anti-ferromagnetic ordering in 2D systems~\cite{Mazurenko2017}, hidden anti-ferromagnetic correlations in 1D Heisenberg spin-chains~\cite{hilkerRevealingHiddenAntiferromagnetic2017}, and have been central to the continued development of our understanding of intriguing phenomena such as superconductivity at high temperatures~\cite{chiuStringPatternsDoped2019,koepsellImagingMagneticPolarons2019}. 

With the exception of two studies on bilayer Mott insulators~\cite{preissQuantumGasMicroscopy2015,koepsellRobustBilayerChargePumping2020}, investigations with QGMs have been constrained to physics in one or two spatial dimensions due to imaging limitations. Full $3$D tomography has only been demonstrated in large-spacing lattices~\cite{Nelson2007} and in $3$D arrays of optical tweezers~\cite{barredoSyntheticThreedimensionalAtomic2018}, where the spacing between atom layers is an order of magnitude larger than that of the system reported on here.

In this article, we present a novel method for the tomographic detection of ultracold $^{87}$Rb atoms sparsely populating a cubic optical lattice potential with a lattice spacing of \SI{532}{\nano\meter} in all directions. Our experimental apparatus is a QGM of the type presented in  Refs.~\cite{Bakr2009,shersonSingleatomresolvedFluorescenceImaging2010}. A schematic of the experiment is shown in Fig.~\ref{fig:One}. The atoms are placed within the (dynamically translatable) focus of a microscope objective with a numerical aperture (NA) of $0.69$. By pinning the atoms in a deep optical lattice, we can use high-contrast fluorescence imaging to determine the atoms' position to within a single site of the lattice.

The microscope objective is mounted on a piezo-driven closed-loop objective scanner with a scan range of \SI{400}{\micro\meter}, which enables accurate control of its vertical position to nanometer precision. It is this vertical control that allows us to perform tomographic scans. Within a single realization of the experiment, we acquire multiple fluorescence exposures, where the position of the microscope objective is translated between images. Thus, we can extract an atom’s position not only in the two horizontal dimensions but also along the (vertical) line-of-sight. The methodology is similar to the one discussed in Ref.~\cite{Nelson2007}. However, in the presented work the distance between adjacent lattice planes is an order of magnitude smaller. As a result, the atoms in neighbouring planes do not simply contribute to an inhomogeneous background~\cite{Nelson2007,koepsellRobustBilayerChargePumping2020}, and the knowledge of the axial behaviour of the imaging system's point spread function (PSF) must be used to determine the atoms' position along the line of sight.
Here lies the novelty of our work in the context of QGM studies, and a successful application will enable spatially resolved studies of many-body systems in three dimensions.

\section{Experimental procedure}

\begin{figure}[tp]
  \centering
    \includegraphics[width=1\columnwidth]{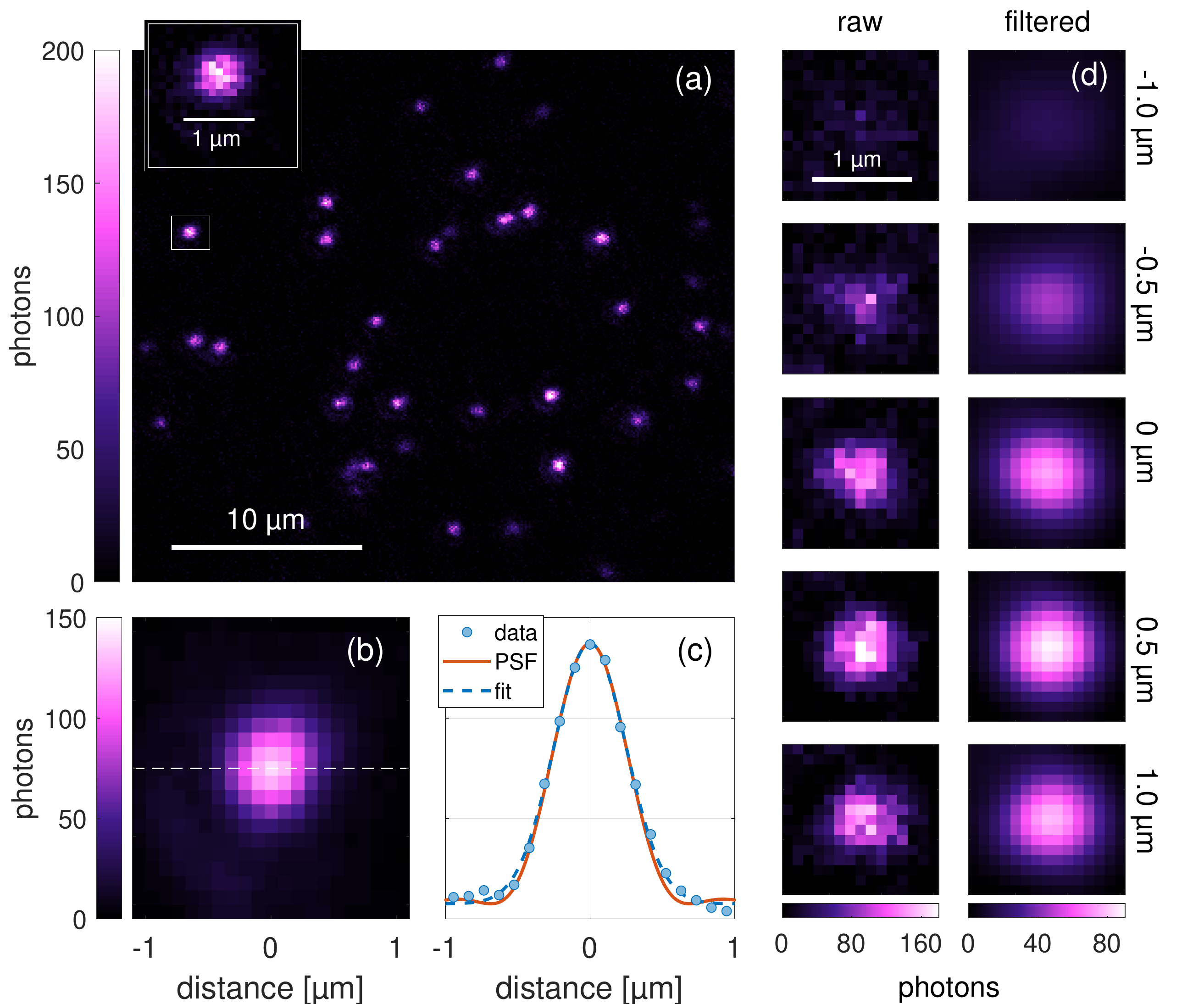} 
  \caption{Fluorescence signals from single atoms. (a) An image of about $40$ atoms. From the atom in the inset we collect approximately $8000$ photons in total. (b) The $20$ brightest signals in (a) are superimposed to create an averaged image. (c) A horizontal cut through the averaged image is fit by a Gaussian function (blue, dashed). The simulated PSF as provided the manufacturer is also plotted (orange, solid). (d) A series of five fluorescence images of a single atom, where the microscope objective was translated by \SI{0.5}{\micro\meter} between consecutive shots. (left column) Raw data. (right column) Data treated with a low-pass filter. The numbers to the right mark the position of the objective, referenced to the center position of the series of images.} 
  \label{fig:Two}
\end{figure}

The full experimental sequence proceeds as follows: initially, atoms are loaded in a magneto-optical trap and subsequently transferred to a magnetic quadrupole potential. The atoms are then cooled by forced evaporative cooling methods as detailed in Refs.~\cite{Bason2018, Heck2018, Eliasson2019} and held in a crossed optical dipole trap at a wavelength of \SI{1064}{\nano\meter}. This results in a cold cloud of about $10^6$ atoms at an approximate temperature of \SI{800}{\nano\kelvin}. The atoms are then loaded into a \SI{940}{\nano\meter} optical tweezer potential generated by directly imaging the pattern displayed on a digital-micromirror device (DMD) through the microscope objective onto the atoms. The resulting tweezer has a $1/e^2$ waist of \SI{780(20)}{\nano\meter}. To sparsely load atoms into the lattice, the tweezer depth is set to \SI{20}{\nano\kelvin}, such that the atoms are merely levitated against gravity. The atoms are then loaded into the vertical lattice and allowed to diffuse within their respective lattice planes. Subsequently, the two horizontal lattice axes are ramped on, and the atoms are frozen in a cubic lattice with a depth of $2000\,E_\mathrm{r}$, where $E_\mathrm{r} = h^2/2m\lambda^2$ is the recoil energy, $m$ is the mass of the atom, and $\lambda = \SI{1064}{\nano\meter}$ is the wavelength of the lattice light. Finally, the atoms are exposed to molasses light configured as in Ref.~\cite{shersonSingleatomresolvedFluorescenceImaging2010}, and their fluorescence is captured by the objective and recorded on an  electron-multiplying CCD camera. A typical experimental image is shown in Fig.~\ref{fig:Two}(a).

\begin{figure*}[htp]
  \centering
    \includegraphics[width=1\textwidth]{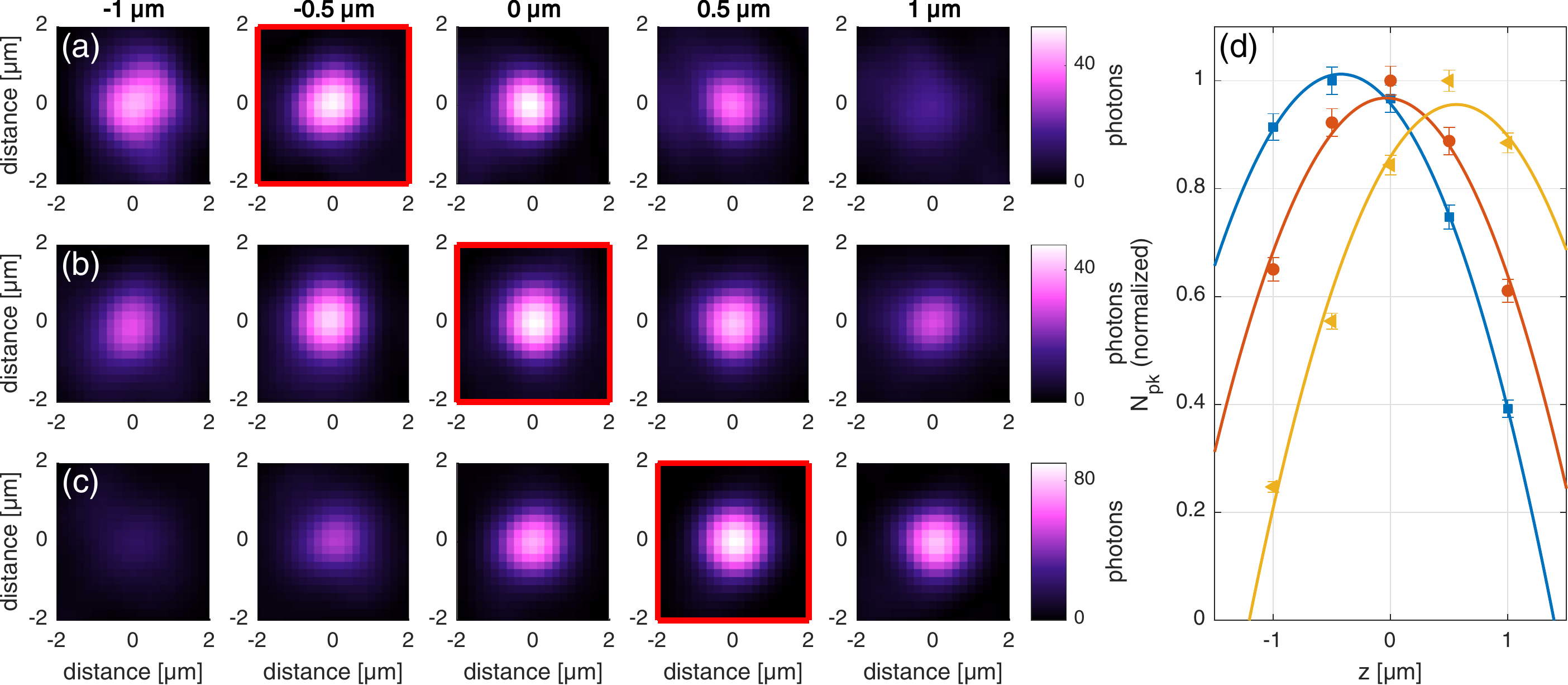} 
  \caption{Determination of an atom's vertical lattice plane. (a)-(c) Fluorescence images of the atoms loaded in the three different planes. (d) Measured normalized peak intensity plotted as a function of the relative position $z$ and corresponding parabolic fits. The three traces corresponds to atoms loaded in the $k=-1$ (blue squares), $k=0$ (red circles) and $k=1$ (yellow triangles). The errorbars represent the shot noise and have been normalized by the peak count.} 
  \label{fig:Three}
\end{figure*}

To identify atoms, a peak-finding algorithm is applied to all the images. In order to determine the resolution of our imaging system, we superimpose $20$ of the brightest signals from the image shown in Fig.~\ref{fig:Two}(a), where the atoms' centers are determined by a $2$D Gaussian fit applied subsequent to the peak finding. In this way we create an averaged image of an atom, as shown in Fig.~\ref{fig:Two}(b). A horizontal cut through the image is fit with a $1$D Gaussian as shown in Fig.~\ref{fig:Two}(c) and used to determine the resolution limit given by the Rayleigh criterion, resulting in $r_\textrm{min} = \SI{713(14)}{\nano\meter}$, in accordance with the manufacturer's specifications.

\section{\label{sec:tomo} Tomographic detection}

In this section, we present a proof-of-principle experiment demonstrating how the position of an atom trapped in our optical lattices can be determined to within a single lattice site in all three dimensions as shown in Fig.~\ref{fig:Three}.

In order to tomographically reconstruct the position of an atom along the vertical lattice planes, we measure its intensity as a function of the objective position. To obtain the data presented here, the atoms were imaged in five \SI{200}{\milli\second} long exposures separated by the same amount of time. During the separation time, the objective was translated through a set of positions $\{z_k\}$, where $k = \{-2,\ldots,2\}$, spanning a total range of \SI{2}{\micro\meter}. Figure~\ref{fig:Two}(d) shows raw and low-pass filtered images of the same atom obtained in this manner. The light intensity scattered by the atom increases up until the fourth image, where the atom is in focus, and then decreases in the fifth image as the atom moves out of focus. This information can be used to pinpoint into which plane along the line-of-sight the particular atom was loaded. As a measure of this effect, we use the peak count $N_\mathrm{pk}$ which is the sum of the photon counts from within a circle with a diameter of 7 pixels around the center of the atom (corresponding to the resolution limit of our imaging system).

For each static atom, we calculate $N_\mathrm{pk}$ using the filtered image as a function of the objective position. The expected axial intensity $I_\textrm{PSF}$ according to the PSF for an ideal imaging system is~\cite{guAdvancedOpticalImaging1999}
\begin{align}
I_\textrm{PSF}(z) = I_0\,\mathrm{sinc}^2(\xi z),\label{eq:intensity_axial}
\end{align}
where $I_0$ is the peak intensity and $\xi = \frac{\pi}{2 \lambda}\,\textrm{NA}^2$. A Taylor expansion of $I_\textrm{PSF}$ around the origin yields an inverse parabola.

The experiments are subject to moderate heating effects. The resulting atom loss (and thermal hopping) during the imaging procedure mainly stems from one of the axes of the optical lattice and the fact that our molasses beams do not cover the entire atom cloud. Therefore, we lose approximately \SI{25}{\percent} of our atoms over the course of the experiment, and of the remaining atoms, about \SI{10}{\percent} are detected in the same horizontal location throughout the five images. As a result, we constrain our analysis to a \SI{20}{\micro\meter}-by-\SI{20}{\micro\meter} region-of-interest where our molasses beams overlap. Additionally, we only consider atoms that remain static in all five images (to within the precision of a single lattice site) for the total \SI{1.0}{\second} duration of the imaging procedure.

We provide representative traces of atoms trapped in the $k = -1, 0,$ and $+1$ planes in Fig.~\ref{fig:Three}(a-c). Each measured set of $N_\mathrm{pk}$ is fitted with an inverse parabola as displayed in Fig.~\ref{fig:Three}(d), where the curvature of the fit is limited to the theoretical value given by the Taylor expansion of equation~\eqref{eq:intensity_axial}. From the quadratic fit, the vertical positions of the atoms are determined as marked with the red square in Fig.~\ref{fig:Three}(a)-(c). For a system with limited loss, this method can be extended to provide a complete tomographic reconstruction of single atom positions in a sparsely loaded optical lattice. Extensions beyond sparse loading are discussed in Sec.~\ref{sec:sims}.

Our method requires that the objective scans through the vertical lattice plane of where the atom is situated. For atomic clouds distributed over a larger extent along the vertical direction more images are needed. More images or longer exposures will increase the accuracy of locating the atoms, which will be explored further in future experiments. The absolute limit, however, is set by the lifetime of the atoms in the lattice and the efficiency of the molasses cooling used.

 In future work, we will focus on realizing tomographic reconstruction of all imaged atoms by mitigating heating effects (e.g. via improved lattice laser stabilization and the improved homogeneity of the molasses) and improving our atom detection methods and imaging SNR, including a thorough analysis of the relative stability of the microscope and the vertical lattice.

\section{\label{sec:sims} Limits and extensions of our method}
In this section, we examine the limits of our tomographic approach and its extension in two physically distinct cases. The first case is that of a sparsely filled lattice, and the second considers a Mott insulator at near-unity filling, that is, with only few vacancies. 
Both cases are investigated by simulating fluorescence images which are generated by propagating the theoretical PSF of the imaging system along its line of sight using the Rayleigh-Sommerfeld transfer function. Images are generated for a series of lattice planes $\{ z_k \}$, separated by \SI{532}{\nano\meter}. To obtain images with multiple atoms, we superimpose propagated PSFs for each individual atom in a given configuration.
For a system where the heating of atoms in the optical lattice is minimized, additional exposures during the experimental sequence are possible. Here we assume that seven images can be taken in each experimental sequence. Furthermore, we assume that the lattice site positions are known.

In an experiment, the number of photons captured from an atom will vary from atom-to-atom e.g.\ due to spatial inhomogeneities in the cooling light. This is included in the simulation by adding normally-distributed fluctuations of \SI{10}{\percent} for each atom. Shot noise is also added to the images, where we assume a mean photon count of \num{2e4} from each atom, as has previously been achieved in a similar system~\cite{Bakr2009}.

\subsection{\label{subsec:sparse} Sparse loading}

First, we consider the limit of sparse lattice loading. In QGM experiments, atoms are distinguishable transversely due to the high-NA imaging system, but we consider here whether or not different atom configurations along the line-of-sight can be distinguished using our tomographic method.

We create nine possible configurations of 2--5 atoms stacked in different lattice planes. The atoms are distributed across 5 planes $k = -2, ..., 2$. A total of seven images are created for the planes  $k = -2, ..., 4$, where the lattice planes $3$ and $4$ are always vacant. The chosen asymmetry will become clear shortly.

\begin{figure*}[tp]
  \centering
    \includegraphics[width=1\textwidth]{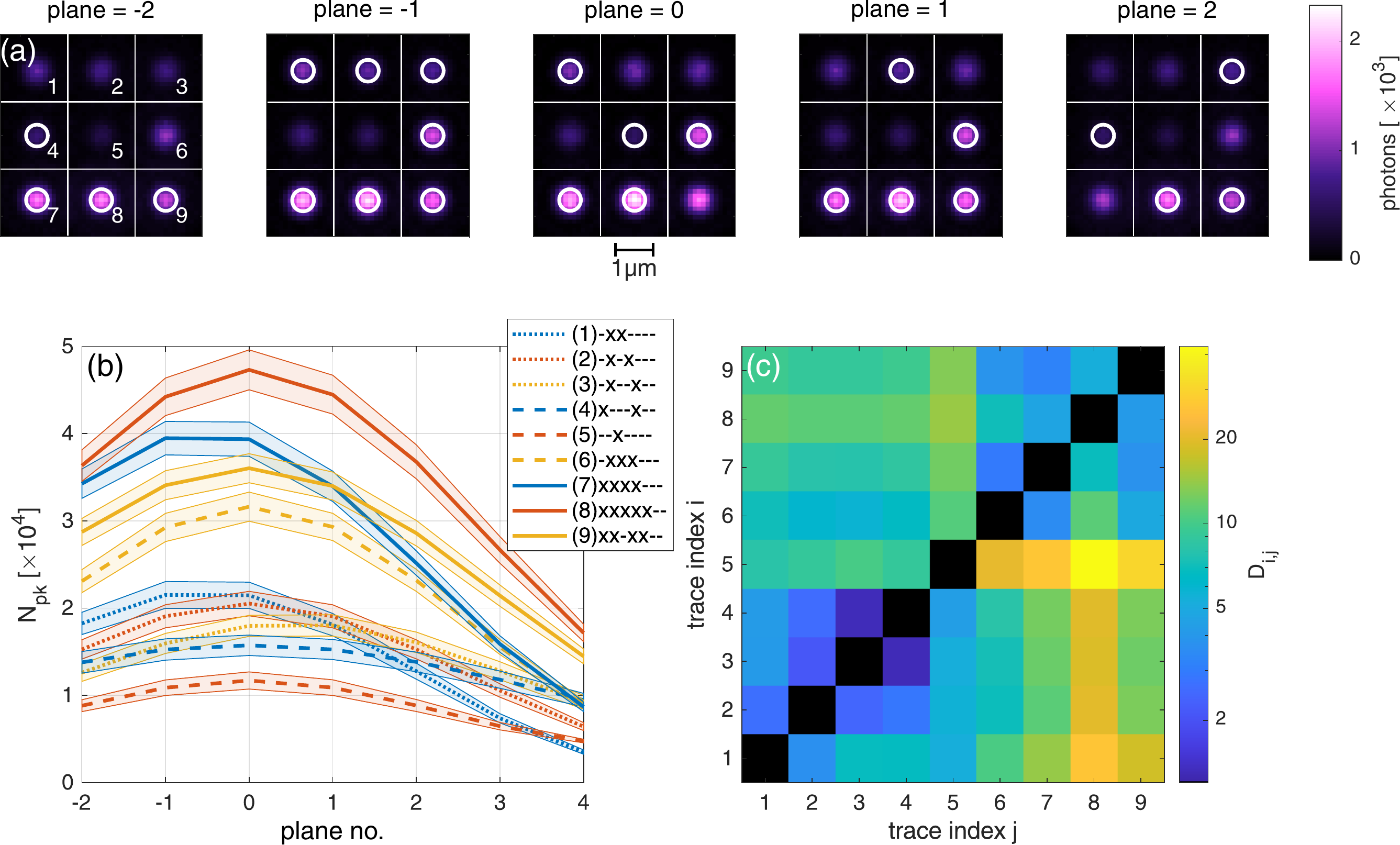} 
  \caption{Simulation of stacked atoms. (a) Simulated images of atoms placed in an optical lattice with \SI{10}{\percent} fluctuations in the collected photon number from the individual atoms and shot noise. The white circles denote the atoms' position. (b) The peak intensity plotted as a function of the lattice planes with the shaded area indicating the standard deviation of repeated realizations with fluctuations present. In the legend, the positions of atoms and vacancies are marked with ``x'' and ``-'', respectively. (c) The largest distance between the traces in terms of the standard deviation shown on a logarithmic scale. The index number is defined in the first frame of (a).} 
  \label{fig:Four}
\end{figure*}

Figure~\ref{fig:Four}(a) shows the images of the non-vacant lattice planes with the atom configurations denoted as 1-9. The atoms' true positions are marked with a white circle on the image.
As an example configuration $6$ has an atom is present in the planes $k = -1, 0,$ and $1$, while the other planes are vacant. We denote this configuration by (${\mathrm{-}\mathrm{x}\,\mathrm{x}\,\mathrm{x}\mathrm{-}\mathrm{-}\mathrm{-}}$)

To acquire statistics due to the two modelled noise sources mentioned above, multiple realizations of the same atom configuration are generated, and from these we extract the trace of $N_\mathrm{pk}$ as a function of lattice plane. Figure~\ref{fig:Four}(b) shows the average of the traces where atoms can be found, and the shaded area denotes the standard deviation. We see that the traces differ more when imaged further away from focus. A higher distinguishability is achieved by increasing the number of images in this region and hence the chosen asymmetry of the planes imaged.

To quantify the similarity of the traces, we determine the distance between two traces $i$ and $j$ in plane $k$, $\Delta_{i,j,k}=|N_{_\mathrm{pk},i,k}-N_{_\mathrm{pk},j,k}|$. The distance is calculated in terms of the standard deviation of the peak count $\sigma_{i,k}$ for trace $i$, with the mean value given as
\begin{equation}
    {D}_{i,j} = \frac{1}{N_\mathrm{planes}}  \sum_k\frac{\Delta_{i,j,k}}{\sigma_{i,k}}.
\end{equation}

The distances between each of the traces from Fig.~\ref{fig:Four}(a) are shown in Fig.~\ref{fig:Four}(c). Here we see that the configurations ($3$) (${\mathrm{-}\mathrm{x}\mathrm{-}\mathrm{-}\mathrm{x}\mathrm{-}\mathrm{-}}$) and ($4$) (${\mathrm{x}\mathrm{-}\mathrm{-}\mathrm{-}\mathrm{x}\mathrm{-}\mathrm{-}}$) are the closest with ${D}_{3,4} = 1.3$ with $\mathrm{max}(\Delta_{3,4,k}/\sigma_{3,k})=2.2$. This underscores the fact that while two atom traces may overlap in certain planes, there always exists a region where the $\Delta_{i,j,k}$ is sufficiently large, such that any ambiguity can be resolved. Thus, all traces shown in Fig.~\ref{fig:Four}(b) can be distinguished in an experiment with sufficiently mitigated loss and heating. Additionally, to relax the requirements on the number of exposures needed per experimental run, one can take images at a subset of the planes considered here or make use of the advanced imaging techniques considered briefly in Sec.~\ref{sec:outlook}.

\subsection{Finding holes in a Mott-insulating state}

Now we shift our focus to the case of a Mott-insulating state with a filling fraction close to unity. With few vacancies in the system, the dip in photon counts due to these vacancies can be used to determine the atomic distribution. In the following, we discuss a simple extension of our tomographic method for such vacancy detection and determine its efficiency.

\begin{figure*}[tp]
  \centering
    \includegraphics[width=1\textwidth]{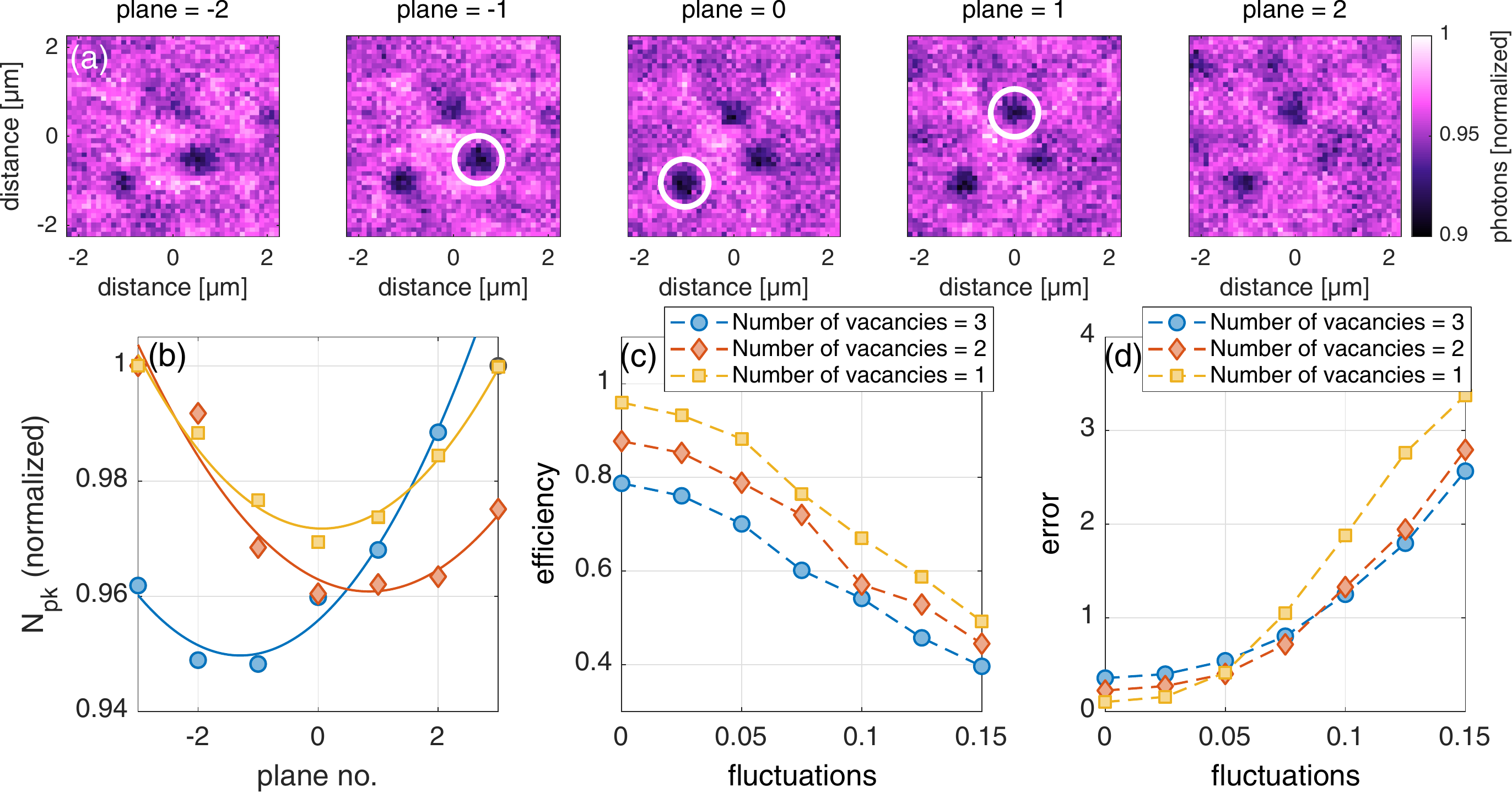}
  \caption{ Simulation of a three-dimensional Mott-insulating state close to unity filling. (a) The images show the central $5\times5\times5$ region of a $17\times17\times17$ Mott insulator. The vacancies are marked with white circles. (b) The traces of the peak count for the vacancies in (a) plotted as a function of the vertical lattice plane.  (c) The efficiency and (d) error of the vacancy detection investigated as a function of the fluctuations in the collected photon number for different number of vacancies.} 
  \label{fig:Six}
\end{figure*}

Figure~\ref{fig:Six}(a) shows the simulated images with three vacant sites in different planes. The vacant positions are limited to a $5\times5\times5$ lattice centered within a $17\times17\times17$ unity filled lattice to simulate the background signal of a large insulating state. As edge effects are present due to the finite size of the Mott insulator, the images are normalized to a background created by averaging 400 realizations with different vacancy positions.

In this case, we analyze seven images generated symmetrically around the five planes in which vacancies can be present.

First, the images are filtered by a low pass filter to limit the effect of shot noise. They are then deconvolved with the PSF of the system using the Lucy-Richardson method~\cite{RichardsonBayesianBasedIterativeMethod1972} which increases the contrast of the images. A minimum-finding algorithm is then used to localize the vacancies in each deconvolved image.

In order to determine the lattice plane of the vacancy, we analyze the $N_\mathrm{pk}$ peak count as a function of the vertical lattice plane $k$, as in Sec.~\ref{sec:tomo}. The vacancy position is then determined by applying a quadratic fit and extracting the position of its minimum. In this case, the curvature of the parabola is positive as the vacant site gives rise to lower photon counts. To limit the fit, the curvature is bounded such that it cannot vary by more than \SI{50}{\percent} from the theoretical predicted curvature from the expansion of equation \eqref{eq:intensity_axial}. Since this Taylor expansion is only valid close to the extremum, the fit is applied to $\pm3$ planes centered around the minimum of the trace, and any points outside of this range are ignored. That is, some fits are done to fewer than seven points, but all fits use at least four points. 

As above, shot noise and fluctuations, normally distributed around the mean photon number collected from each atom, are added to the images. The tomographic method is then applied to images generated at different levels of fluctuations.
To quantify the quality of the method, we look at the efficiency and the error of vacancy detection. The efficiency is taken to be the probability of detecting and locating a vacancy correctly. The error gives the number of false positives (when the method detects a vacancy at a position where there is none). This manifests itself both as the detection of a false vacancy due to the noise or an incorrectly located hole. Figure~\ref{fig:Six}(c) shows how the efficiency of the method decays at large fluctuation levels, mainly due to the quadratic fit failing to locate the vacancy in the correct lattice plane. The error shown in Fig.~\ref{fig:Six}(d) arises initially due to this same incorrect plane determination, but as the fluctuations rise above \SI{5}{\percent}, the peak-finding algorithm also starts to mistakenly detect false vacancies in the noise. It should also be noted that the efficiency of the method declines when inserting more holes into the system as atoms next to or on top of each other are hard to separate and locate.

\section{\label{sec:outlook}Conclusion and Outlook}

In this manuscript, we have experimentally demonstrated a tomographic approach to determining an atoms' position within a three-dimensional cubic lattice, given that the lattice is sparsely loaded. In addition, simulations suggest that similar methods can be used to detect holes in a unity-filled, three-dimensional Mott insulating state.

While the effectiveness of these methods in the intermediate regime (e.g.\ \SI{50}{\percent} filling) is left for future study, their utility in both of the investigated regimes opens up interesting possibilities for further studies. Additionally, improved imaging and image analysis techniques will likely expand the regimes in which our methods are applicable and useful. These techniques, like the tomographic methods discussed in Ref.~\cite{albertiSuperresolutionMicroscopySingle2016a}, draw largely on existing biological, geo- and astrophysical methods. Such possibilities include methods based on principal component analysis~\cite{liAtmosphericTurbulenceDegradedImage2007}, more advanced deconvolutional~\cite{jefferiesRestorationAstronomicalImages1993, mcnallyThreeDimensionalImagingDeconvolution1999} and optical coherence tomographic methods~\cite{woolliamsSpatiallyDeconvolvedOptical2010}. One could even use a spatial light modulator to shape the PSF of the microscope in the longitudinal plane, as in Ref.~\cite{pavaniThreedimensionalSinglemoleculeFluorescence2009} to allow for easier recognition of an atom's lattice plane.  Methods like these could enable the determination of an atom's three-dimensional location with fewer images, thus relaxing the requirements of long imaging times.

Our sparsely-filled tomographic technique could be applied to study a broad class of transport problems in three dimensions, including spintronics~\cite{langbeinTimeOptimalInformation2015}, light-harvesting systems~\cite{chinNoiseassistedEnergyTransfer2010,tomasiCoherentControllableEnhancement2019,mohseniEnvironmentassistedQuantumWalks2008}, or the spreading of impurities in a spin system~\cite{Fukuhara2013, hildFarfromEquilibriumSpinTransport2014}. One particular experiment that fulfills the condition of sparse loading is the realization of quantum random walks in $3$D, which has been investigated before in $1$D in QGMs~\cite{preissStronglyCorrelatedQuantum2015, karskiQuantumWalkPosition2009}, solid-state qubits~\cite{yanStronglyCorrelatedQuantum2019} and in photonic systems~\cite{peretsRealizationQuantumWalks2008}, among others. To study interesting dynamics in this setting, one could, for example, tilt the lattice (which is already a non-trivial classical problem~\cite{evstigneevDiffusionColloidalParticles2008}) or modulate it to study the diffusion of higher motional states.

In the bulk, our method allows for three-dimensional quantum simulations that were hitherto inaccessible, for example, a more precise determination of the superfluid-to-Mott-insulator transition in three dimensions \cite{prosniakCriticalPointsThreedimensional2019}.
Such bulk methods also enable the study of dynamics such as many-body entanglement~\cite{lukinProbingEntanglementManybody2019}, localization~\cite{choiExploringManybodyLocalization2016, rubio-abadalManyBodyDelocalizationPresence2019}, and quantum walks of defects in Mott-insulating states~\cite{mondalQuantumWalksInteracting2020}. In this new regime of quantum simulation, the dimensionality of the system renders numerical computation of the system difficult~\cite{eisertColloquiumAreaLaws2010}.

\begin{acknowledgments}
 The authors would like to thank Plamen Petkov for creating Fig.~\ref{fig:One}, and acknowledge financial support from the European Research Council, H2020 grant 639560 (MECTRL);
 the Lundbeck Foundation, the Carlsberg Foundation, and the Villum Foundation.
\end{acknowledgments}

\end{document}